\newcommand{\gapr}{\raisebox{-.6ex}{\mbox{
$\stackrel{>}{\mbox{\scriptsize$\sim$}}\:$}}}
\newcommand{\lapr}{\raisebox{-.6ex}{\mbox{
$\stackrel{<}{\mbox{\scriptsize$\sim$}}\:$}}}

\documentclass{aa}
\usepackage{graphicx}
\begin{document}
\title{The Chandra LETGS high resolution X-ray spectrum of 
the isolated neutron star RX\,J1856.5--3754}

\author{V. Burwitz\inst{1} \and V.~E. Zavlin\inst{1} 
\and R. Neuh\"auser\inst{1} \and P. Predehl\inst{1} \and J. Tr\"umper\inst{1}
\and A.~C. Brinkman\inst{2}}

\institute{Max-Planck-Institut f\"ur extraterrestrische Physik, 
P.O. Box 1312, D-85741 Garching, Germany
\and Space Research Organization of the Netherlands, 
Soorbonnelaan 2, 3584 CA Utrecht, Netherlands}

\offprints{V. Burwitz, {\tt burwitz@mpe.mpg.de}}

\date{Received August 1, 2001; accepted September 18, 2001 }

%

\abstract{
We present the {\it Chandra} LETGS X-ray spectrum of the 
nearby ($\simeq 60$\,pc) neutron star RXJ1856.5--3754.
Detailed spectral analysis of the combined X-ray and 
optical data rules out the nonmagnetic neutron star 
atmosphere models with 
hydrogen, helium, iron and solar
compositions.
We 
also conclude
that strongly magnetized atmosphere models 
are unable to
represent the data.
The data can be explained with a two-component blackbody 
model. 
The harder component with temperature of 
$kT_{\rm bb}^\infty \simeq 63$\,eV and a radius 
$R_{\rm bb}^\infty\simeq 2.2$\,km of the emitting region 
well fits the X-ray data and can be interpreted as 
radiation from a hot region on the 
star's 
surface.
\keywords{Stars: atmospheres -- stars: individual (RX~J1856.5--3754):
-- stars: neutron -- X-rays: stars}
}
\titlerunning{The $Chandra$ LETGS spectrum of
RX\,J1856.5--3754}
\authorrunning{Burwitz et. al.}
\maketitle

\section{The neutron star RXJ1856.5--3754}
It was suggested by Walter et al. (\cite{Waletal96}) and 
supported by Neuh\"auser et al~(\cite{Neuetal97}) that 
RXJ1856.5--3754 (or RXJ1856 for short) is an isolated 
neutron star (NS).
It has a strong and non-variable X-ray flux
$\simeq 1.5\times 10^{-11}$~erg~s$^{-1}$~cm$^{-2}$
and a soft spectrum with a blackbody temperature 
$kT_{\rm bb}^\infty\simeq 57$~eV.
Thousands of such NSs were expected in RASS (e.\,g., 
Colpi et al.~\cite{Coletal93}; Madau \& Blaes \cite{MadBla94}),
moving fast through the interstellar medium.
Only a few such objects have been found, three of them show 
pulsations on time-scale of seconds (Haberl et al.~\cite{Habetal97}, 
\cite{Habetal99}; Hambaryan et al.~\cite{Hametal01}).
Neuh\"auser \& Tr\"umper (\cite{NeuTru99}) argued that the 
number of isolated NSs expected in RASS was overestimated 
mainly due to unrealistic velocity distributions.

Walter \& Matthews~(\cite{WalMat97}) and Neuh\"auser et 
al.~(\cite{Neuetal98}) found an optical counterpart for 
RXJ1856 with $V$\,$\simeq$\,26\,mag.
This and the large proper motion of $\simeq 0.33$\,mas\,yr$^{-1}$
(Walter~\cite{Wal01}; Neuh\"auser~\cite{Neu01})
are additional arguments that it is indeed an isolated NS.
Walter (\cite{Wal01}) also detected parallactic motion,
determined the distance to the source $d=61^{+9}_{-8}$\,pc 
and suggested that RXJ1856 could be the remnant of the original
primary of the $\zeta$ Oph system.
This implies a NS age of $\sim 10^6$\,yr.

First spectral modelling of RXJ1856 based on its {\it ROSAT} 
data was presented by Pavlov et al.~(\cite{Pavetal96}; 
hereafter P96), who showed that the optical/$UV$ flux 
predicted by NS atmosphere models depends drastically on the
surface chemical composition.
More recently Pons et al.~(\cite{Ponetal02}; hereafter P02)  
have analyzed combined optical and X-ray data and concluded 
that it may have either a Fe or Si-ash atmosphere.
However, the inferred NS radius $R\,\approx\,6$\,km and mass
$M\,\approx\,0.9\,M_{\odot}$ are not allowed for any plausible
equation of state of the NS inner matter.

RXJ1856 was observed with the $Chandra$ Low Energy Transmission 
Grating Spectrometer (LETGS; Brinkman et al.~\cite{Brietal00}).
First preliminary results were presented by Burwitz et 
al.~(\cite{Buretal01}).
Here we describe the LETGS data (\S2), results of spectral (\S3)
and timing (\S4) analysis and discuss implications on the nature 
of RXJ1856 (\S5).
%

\section{LETGS data extraction}

\begin{figure}
   \centering
   \includegraphics[angle=90,width=8.8cm]{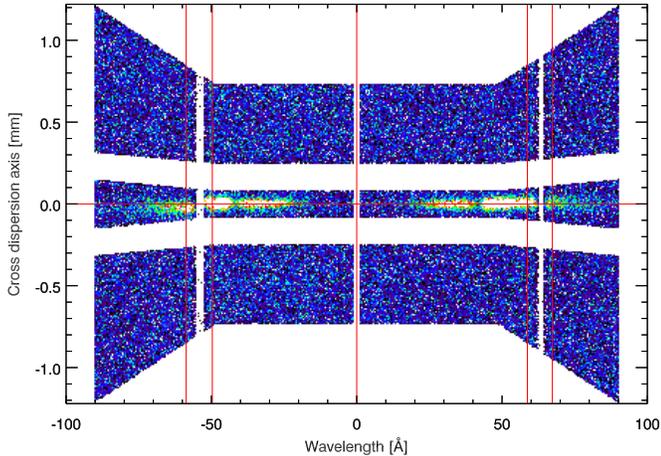}
   \caption{Regions used for extracting the source and background 
            dispersed spectra. The vertical red lines on either
	    side of the gaps between the HRC-S detector plates indicate the 
	    the areas that are masked out avoid getting artificial spectral 
	    features due to the dithering across the plate edges (see POG).
        \label{ExtrReg}}
    \end{figure}

RXJ1856 was observed on March 10, 2000 with the standard 
LETGS (LETG + HRC-S) configuration in a 56.1\,ks exposure. 
The LETGS spectrum of RXJ1856 was extracted from the 
level 1.5 event file.
The only pulse-height filtering applied to the data was the 
removal of photons with a pulse-height amplitude equal to 255
as other pulse-height filters do not reduce the background level
evenly for all energies. 
We used the extraction region recommended in the $Chandra$ 
Proposers' Observatory Guide\footnote{
{\tt http://asc.harvard.edu/udocs/docs/docs.html}} [POG])
for extracting the source spectrum (see Fig.~\ref{ExtrReg}).
For the background, large regions above and below the
source extraction area are selected.
Towards longer wavelengths $\lambda $\,$>$\,48.4\,\AA\ the 
background regions become wider in order to maintain the 
constant ratio of areas between source and background spectral
bins (here the ratio is equal to 6). 
The measured dispersed source count rate is 290\,$\pm$\,6\,ks$^{-1}$
in the 0.15--0.82~keV range (where the source spectrum prevails
over the background).
The extracted source spectrum (binned in 686 spectral bins) 
and the most up-to-date effective area tables for the 1st 
order (status of October 31, 2000)\footnote{{\tt 
http://asc.harvard.edu/cal/Links/Letg/User}}
were used for spectral fits.

%

\section{Spectral analysis}

To search for spectral lines, we used the method applied by 
Pavlov et al.~(\cite{Pavetal01}) to the dispersed LETGS data
on the Vela pulsar. 
We first binned the extracted source-plus-background and 
background spectra in 0.02\,\AA~ bins. 
Then we grouped 10--20 sequent bins and estimated the 
deviation of source counts in each bin of a given group 
from the mean value in the group.
This method revealed maximum deviation at a 2.9\,$\sigma$ level. 

Since there exist the $ROSAT$ PSPC data ($\sim 20,200$ counts 
collected in a 6.3\,ks exposure), we first checked whether the 
LETGS and PSPC data yield consistent results in spectral fits. 
The blackbody fits showed that the allowed domains of the 
fitting parameters do not formally overlap with each other:
$kT_{\rm bb}^\infty = 57\pm 3$~eV, 
$R_{\rm bb}^\infty=(3.7\pm 0.7)\, d_{60}$~km ($d_{60}=d/60$~pc) 
and $n_{\rm H,20}=n_{\rm H}/(10^{20}~{\rm cm}^{-2})=1.47\pm 0.25$
(the uncertainties are given at a 3$\sigma$ level)
with $\chi^2_\nu=1.5$ for the PSPC data, and 
$kT_{\rm bb}^\infty = 63\pm 3$~eV,
$R_{\rm bb}^\infty=(2.2\pm 0.3)\, d_{60}$~km and
$n_{\rm H,20}=1.03\pm 0.20$
($\chi^2_\nu=1.0$) for the LETGS data.
%
   \begin{figure}
   \centering
   \includegraphics[angle=0,width=8.8cm]{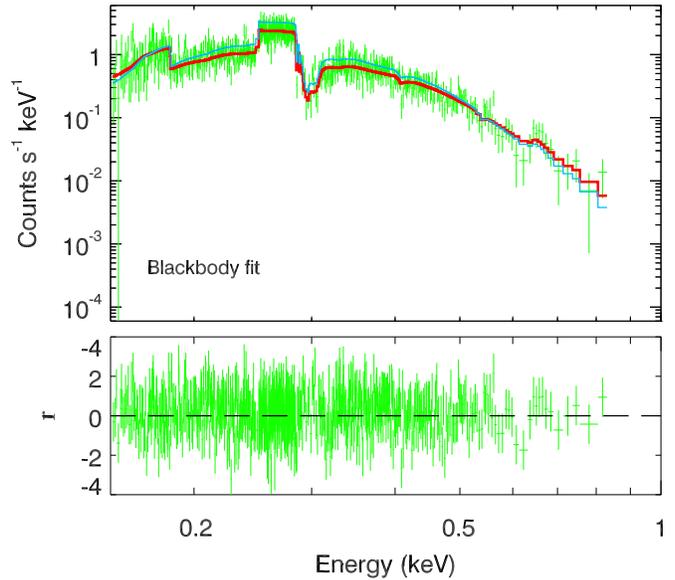}
   \caption{The 
best blackbody fit to the LETGS spectrum (red curve)
with the blackbody model given by the best PSPC fit (blue curve).
              \label{BBMod}}
    \end{figure}
%
   \begin{figure*}
   \centering
   \includegraphics[angle=0,width=15.0cm]{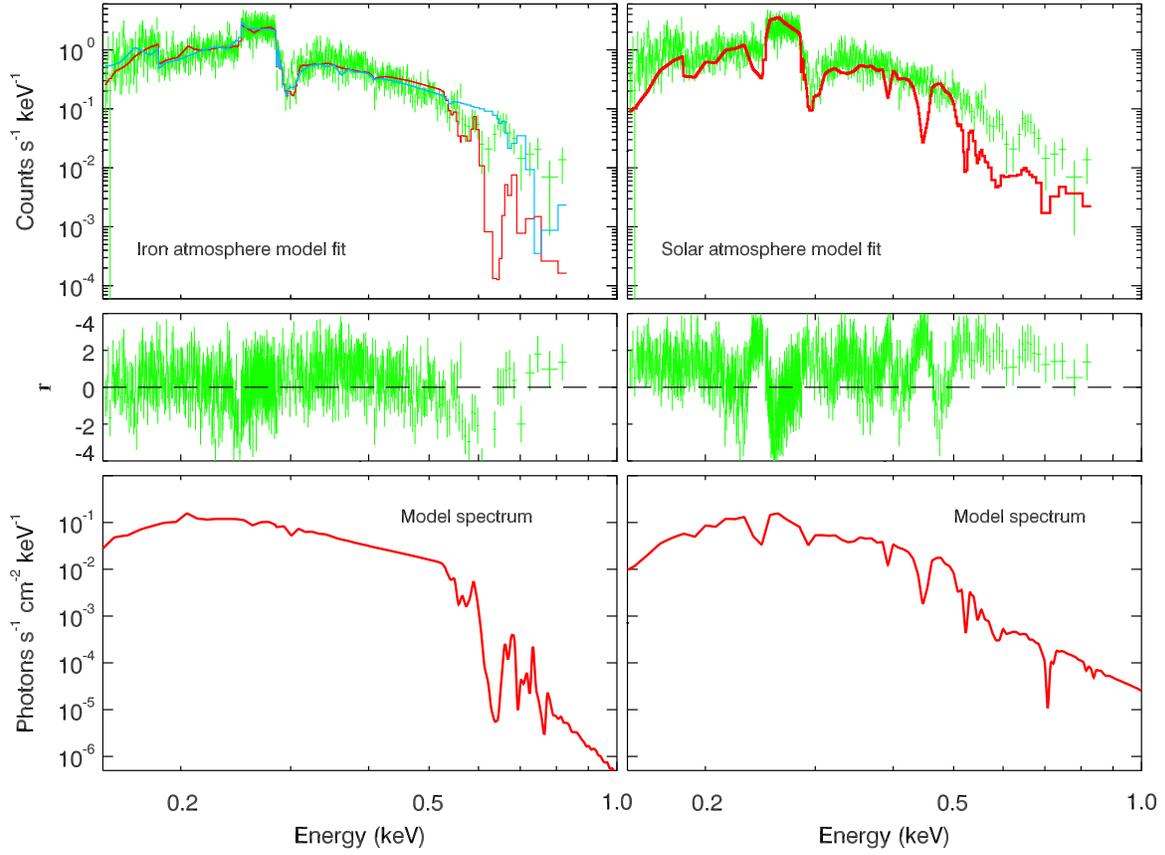}
   \caption{The LETGS spectrum fitted 
with the Fe 
({\it left}) and solar mixture ({\it right})
atmosphere models (see text for details).
              \label{FeSoMod}	}      
    \end{figure*}
The difference between the blackbody models given by
the two fits (see Fig.~\ref{BBMod}) is most likely  
attributed to uncertainties in the calibration of 
LETGS and PSPC.

The next step in analysing the LETGS spectrum is to 
apply NS atmosphere models.
The results of P96 and P02 showed that the light 
element, H and He, nonmagnetic atmosphere models 
can be firmly ruled out because the models {\it (i)}
yield too small distances to RXJ1856, 
$d\simeq(5-6)\,(R/10~{\rm km})$\,pc and
{\it (ii)} overpredict the optical flux by a factor 
of 20--30.
P02 found that heavier element atmosphere models can 
be reconciled with the PSPC and optical data at 
$kT=57.4$\,eV, $R=6.0\, d_{60}$\,km for pure Fe, and 
$kT= 58.7$\,eV, $R=6.0\, d_{60}$\,km for Si-ash compositions,
assuming the gravitational parameter
$g_{\rm r}=[1-R_{\rm g}/R]^{1/2}=0.77$ ($R_{\rm g}=2GM/c^2$ 
is the Schwarzschild radius)\footnote{
For the atmosphere model fits we give parameters as 
measured at the NS surface, $T=g_{\rm r}^{-1} T^\infty$ 
and $R=g_{\rm r} R^\infty$}.
Varying $g_{\rm r}$ resulted in negligible changes in the 
inferred model parameters unless the strong absorption line
at (unredshifted) $E_{\rm true}\simeq 0.79$\,keV (due to 
absorption by \ion{Fe}{xviii} ions) penetrates in lower 
(redshifted) energies $E=g_{\rm r} E_{\rm true} < 0.6$\,keV 
and significantly affects the quality of the PSPC spectral fits.
The effects of gravity on the emergent spectra are too small
to affect the results of the spectral fits.
The fit of the PSPC spectrum with Fe atmosphere models of 
Pavlov \& Zavlin (\cite{PavZav00}) yields $d=54\pm 10$\,pc, 
$kT=45\pm 5\,$eV at $R=10$~km and $M=1.4 M_\odot$, that are
well consistent with those obtained by P02.
Although the LETGS spectrum revealed no statistically 
significant features, it is worthwhile trying the heavy 
element atmosphere models on these data.
To fit the LETGS data, we fixed the distance at $d=60$~pc 
and obtained best fitting parameters $n_{\rm H,20}=1.72$,
$kT=40.3$\,eV, $R=12.0\, d_{60}$\,km at $M=1.4 M_\odot$ 
(see red curves in the left panels of Fig.~\ref{FeSoMod}).
The model fits well at (redshifted) $E\lapr0.5$~keV where 
the atmosphere spectrum shows no strong spectral features. 
However, the situation is drastically different at the 
higher energies because of the strong absorption line at 
$E=0.81 E_{\rm true}\simeq 0.64$\,keV (see above) of a 
characteristic width $\sim 70$\,eV and a weaker absorption
complex at $E\simeq 0.55-0.58$\,keV.
We note that the LETGS spectrum shows a strong 
instrumental feature at $E=0.61-0.63$\,keV (POG).
Despite of a formally small value of $\chi^2_\nu=1.2$ obtained
in this fit, the deviation between the model and LETGS spectra
at $E\gapr 0.5$\,keV are large enough to regard the fit as 
unacceptable.
Increasing $g_{\rm r}$ (i.e. reducing the redshift) does not 
improve the fit because the absorption lines remain within 
the observed energy range (see an example of the spectral 
fit at $M=0.2 M_\odot$, $R=9.8$\,km and $g_{\rm r}=0.97$ in 
the upper left panel of Fig.~\ref{FeSoMod} [blue curve]).

The same strong Fe features are expected to be present in the 
Si-ash model spectra as Fe composes 68\% of the ash (P02).
Therefore, the situation with the Si-ash model fit to the 
LETGS spectrum should be similar to the Fe case.
We also applied atmosphere models with a solar composition 
containing 2\% heavy elements (Grevesse \& Noels~\cite{GreNoe93}).
The model spectra show numerous prominent spectral features 
(mainly due to absorption by Fe, C and O ions) in the whole 
observed energy range (see right panels in Fig.~\ref{FeSoMod},
and also Rajagopal \& Romani [\cite{RajRom96}]) which are 
inconsistent with the LETGS data.

\section{Timing analysis}
For the timing analysis we used 11,606 counts extracted from 
a $2\arcsec$-radius circle centered at the zero-order image
(we did not use the dispersed data as they are strongly 
contaminated by background). 
We had to restrict the search for a periodical signal at the
(upper) frequency $f=40$\,Hz because of the time-tag problem
(see POG), unrecovered in this observation. 
As a result, the HRC timing accuracy in this observation was
about 4--5\,ms (vs. 16\,$\mu$s planned).
We applied the standard $Z^2_1$ (Rayleigh) test 
(Buccheri et al.~\cite{Bucetal83})
as well as the method based on the Bayesian statistics
(Gregory \& Loredo~\cite{GreLor96}; Zavlin et al.~\cite{Zavetal00}; 
Hambaryan et al.~\cite{Hametal01}).
The frequency range $f=10^{-3}-40$\,Hz ($P=25~{\rm ms}-10^3$\,s)
revealed no statistically significant signals (maximum 
significance is at a 2.1$\sigma$ level).
Assuming a sinusoidal signal,
 we put an upper limit $f_{\rm p}<8\%$
on its pulsed fraction. 

\section{Discussion and summary}
The most striking result of our $Chandra$ LETGS
data analysis (extended to the optical range [P96; P02])
is the fact that a simple blackbody model yields a much
better fit to the RXJ1856 spectrum than either light or 
heavy element nonmagnetic atmosphere models.
Note that the heavy element models are inconsistent
with the high-resolution LETG spectrum.
On the other hand, nonmagnetic H/He models do not work
either because they predict to high optical fluxes.
A remedy may be provided if RXJ1856 has a strong magnetic
field as indicated by a bow-shock nebula seen
in H$_\alpha$ optical observations 
(van Kerkwijk \& Kulkarni \cite{KerKul00}). 
Similar to the case with the nonmagnetic models,
the magnetized H atmosphere fits result in a too 
small distance and a too high optical flux (P96).
Magnetized Fe atmosphere spectra at $kT\sim 60$~eV
(Rajagopal et al.~\cite{Rajetal97}) reveal a lot of 
strong absorption features in the 0.1--1.0~keV range 
which should have been resolved with the LETGS spectral
resolution.
Hence, we conclude that the available NS atmosphere models
cannot represent both the X-ray and optical data on RXJ1856.  

Regardless of the chemical composition,
radiation emitted from the surface of a magnetized
NS should exhibit spectral lines of the
electron (at $E_{B{\rm e}}=11.6B_{12}$\,keV) and
proton (at $E_{B{\rm p}}=6.3B_{12}$\,eV) cyclotron resonances
($B_{12}$ is the strength of magnetic field in units of $10^{12}$\,G).
The lack of significant features in the LETGS spectrum (0.15--0.82\,keV)
may appear to exclude the magnetic fields 
$B\simeq (1.3-7.0)\times 10^{10}$\,G
and $B\simeq (0.2-1.3)\times 10^{14}$\,G.
But we note that these restrictions on $B$ 
are not very stringent because these lines
may be rather faint if the surface layers are only weakly ionized.
Besides, nonuniformity of the magnetic field over the
surface should lead to a strong smearing of the lines
(Zavlin et al.~\cite{Zav95}).

Remarkably, the X-ray spectrum of the other isolated NS, 
RX~J0720.4--3125, obtained with {\it XMM-Newton} also 
shows no significant spectral features and is well fitted
with a blackbody model of $kT_{\rm bb}^\infty\simeq 86$\,eV
(Paerels et al.~\cite{Paer01}).

P02 argued that a two-component blackbody model
can be reconciled with both the optical and X-ray data
on RXJ1856.
In this model the hard component of $kT^\infty\simeq 55$~eV 
is emitted from a $\sim 20\%$ fractional area on the NS surface
and fits the X-ray data, whereas the soft component of
$kT^\infty\simeq 20$~eV represents radiation from the 
cool surface and matches the optical data.
The required non-uniform distribution of the surface 
temperature may be due to the strong dependance of thermal
conductivity of the NS crust on the magnetic field at 
$B\gapr 10^{12}$~G (Greenstein \& Hartke~\cite{GreHar83}).
At temperatures $kT\lapr 86$~eV and magnetic fields 
$B\gapr 10^{13}$~G hydrogen (if present on the surface) can be 
in form of polyatomic molecules and/or a condensed liquid
(Lai \& Salpeter~\cite{LaiSal97}). 
Although, to our knowledge, no reliable calculations have 
been done, one may speculate that such a condensed matter
surface emits radiation close to the blackbody spectrum at
a temperature close to that of the surface,
as suggested by Pavlov (\cite{P00})\footnote{{\tt http://online.itp.ucsb.edu/online/neustars00/pavlov}}.
Then, the two-component blackbody interpretation may be 
considered as a simplified model of the thermal radiation 
from such a magnetized and relatively cool NS.

New important information for elucidating the nature of 
RXJ1856 is expected to come from the forthcoming 
50\,ks {\it XMM-Newton} and  450\,ks {\it Chandra}
observations.
They will yield 
an order of magnitude 
more counts
in the dispersed spectra than in the data presented above.
This will allow one to take more advantage of the high 
spectral resolution.
Additionally, 
the new data would enable 
an unambiguous detection of
X-ray pulsations from the
source even if the pulsed fraction is as low as 3\%.
Once the period is found 
in the new data,
it could be 
traced back to the earlier {\it ROSAT} and {\it Chandra} 
data. 
This would give estimates on the period derivative
and, as consequence, magnetic field and age of RXJ1856.
The latter, compared with the estimate derived from
the optical observations, would shed more light
on the nature of this enigmatic source.

\acknowledgements{
We thank Fred Walter, George Pavlov and Frits Paerels 
for intensive discussions.
This research was supported by Deutsches Zentrum f\"ur
Luft- und Raumfahrt grant 50\,OX\,0001.
}

%

\end{document}